\documentclass[
reprint,prx, longbibliography,
superscriptaddress,
 amsmath,amssymb,
 aps,
]{revtex4-2}

\usepackage{graphicx}
\usepackage{dcolumn}
\usepackage{bm}
\usepackage{siunitx}
\usepackage{chemformula}
\usepackage{CJKutf8}

\newcommand{\ket}[1]{\left| #1 \right>} 
\let\vaccent=\v 
\renewcommand{\v}[1]{\ensuremath{\mathbf{#1}}} 
 
\newcommand{\h}{\hat}

\begin{document}
\begin{CJK}{UTF8}{gbsn}

\title{Quantum spin probe of single charge dynamics}

\author{Jonathan C. Marcks}
\affiliation{Pritzker School of Molecular Engineering, University of Chicago, Chicago, IL, 60637, United States}
\affiliation{Center for Molecular Engineering and Materials Science Division, Argonne National Laboratory, Lemont, IL, 60439, United States}

\author{Mykyta Onizhuk}
\affiliation{Department of Chemistry, University of Chicago, Chicago, IL, 60637, United States}
\affiliation{Pritzker School of Molecular Engineering, University of Chicago, Chicago, IL, 60637, United States}

\author{Yu-Xin Wang (王语馨)}
\affiliation{Pritzker School of Molecular Engineering, University of Chicago, Chicago, IL, 60637, United States}
\affiliation{Joint Center for Quantum Information and Computer Science, University of Maryland, College Park, MD 20742, United States}

\author{Yu Jin}
\affiliation{Department of Chemistry, University of Chicago, Chicago, IL, 60637, United States}
\affiliation{Pritzker School of Molecular Engineering, University of Chicago, Chicago, IL, 60637, United States}

\author{Yizhi Zhu}
\affiliation{Pritzker School of Molecular Engineering, University of Chicago, Chicago, IL, 60637, United States}

\author{Benjamin S. Soloway}
\affiliation{Pritzker School of Molecular Engineering, University of Chicago, Chicago, IL, 60637, United States}

\author{Masaya Fukami}
\affiliation{Pritzker School of Molecular Engineering, University of Chicago, Chicago, IL, 60637, United States}

\author{Nazar Delegan}
\affiliation{Center for Molecular Engineering and Materials Science Division, Argonne National Laboratory, Lemont, IL, 60439, United States}
\affiliation{Pritzker School of Molecular Engineering, University of Chicago, Chicago, IL, 60637, United States}

\author{F. Joseph Heremans}
\affiliation{Center for Molecular Engineering and Materials Science Division, Argonne National Laboratory, Lemont, IL, 60439, United States}
\affiliation{Pritzker School of Molecular Engineering, University of Chicago, Chicago, IL, 60637, United States}

\author{Aashish A. Clerk}
\affiliation{Pritzker School of Molecular Engineering, University of Chicago, Chicago, IL, 60637, United States}

\author{Giulia Galli}
\affiliation{Pritzker School of Molecular Engineering, University of Chicago, Chicago, IL, 60637, United States}
\affiliation{Department of Chemistry, University of Chicago, Chicago, IL, 60637, United States}
\affiliation{Center for Molecular Engineering and Materials Science Division, Argonne National Laboratory, Lemont, IL, 60439, United States}

\author{David D. Awschalom}
\affiliation{Pritzker School of Molecular Engineering, University of Chicago, Chicago, IL, 60637, United States}
\affiliation{Department of Physics, University of Chicago, Chicago, IL, 60637, United States}
\affiliation{Center for Molecular Engineering and Materials Science Division, Argonne National Laboratory, Lemont, IL, 60439, United States}

\date{\today}

\begin{abstract}
Electronic defects in semiconductors form the basis for many emerging quantum technologies. Understanding defect spin and charge dynamics in solid state platforms is crucial to developing these building blocks, but many defect centers are difficult to access at the single-particle level due to the lack of sensitive readout techniques. A method for probing optically inactive spin defects would reveal semiconductor physics at the atomic scale and advance the study of new quantum systems. We exploit the intrinsic correlation between the charge and spin states of defect centers to measure defect charge populations and dynamics through the steady-state spin population, read-out at the single-defect level with a nearby optically active qubit. We directly measure ionization and charge relaxation of single dark defects in diamond, effects we do not have access to with traditional coherence-based quantum sensing. These spin resonance-based methods generalize to other solid state defect systems in relevant materials.
\end{abstract}

\maketitle
\end{CJK}

Dopants in semiconductors are a key part of modern technology~\cite{Mccluskey_dopants_2018,Koenraad_single_2011} and emerging quantum applications~\cite{Kane_silicon-based_1998,Awschalom_quantum_2018,Barry_sensitivity_2020,wolfowicz_quantum_2021,Burkard_semiconductor_2023}. In traditional semiconductor devices, a high density of dopants enables current flow and logic operations~\cite{Mccluskey_dopants_2018}. Many electronic dopants also carry spin~\cite{Awschalom_quantum_2018}, and controlled doping of materials such as diamond~\cite{Ohno_engineering_2012}, silicon~\cite{Berkman_millisecond_2023}, and various oxides~\cite{Bassett_quantum_2019,wolfowicz_quantum_2021,Stevenson_erbium-implanted_2022} with deep-level defects has led to the coherent manipulation of spin states in optically active electronic defects~\cite{Awschalom_quantum_2018}. These so-called color centers may interact with nearby optically inactive (dark) spins~\cite{Belthangady_dressed-state_2013,sushkov_magnetic_2014,dwyer_probing_2022}, which generally cause central spins to decohere~\cite{Hanson_coherent_2008,Bauch_decoherence_2020,Marcks_guiding_2023}, but may also be leveraged to advance quantum sensing~\cite{Zheng_preparation_2022,Meriles_quantum_2023} and quantum memories~\cite{Degen_entanglement_2021}. However, control of these color centers, and by extension dark spins, requires optical illumination, which may drive the electronic environment out of equilibrium, leading to dynamic processes such as ionization. A better understanding of the coupling between charge and spin degrees of freedom in semiconductors at the single-defect level is necessary for the development of quantum technologies based on solid state spin systems~\cite{Koenraad_single_2011,wolfowicz_quantum_2021}.

Optical detection of single bright defects is now routine~\cite{Awschalom_quantum_2018}, and spin-dependent ionization has lead to high fidelity readout through spin-to-charge conversion (SCC) techniques~\cite{Shields_efficient_2015,Anderson_five-second_2022}. Optically inactive spins in semiconductors, such as electron spins in GaAs and Si quantum dots, are also readily measured at the single-spin level~\cite{Elzerman_single-shot_2004}. These techniques require either a spin-photon interface or specific device fabrication. A generic feature in all these settings is that the spin state of an electronic system is determined by the electron orbital filling. This intrinsic connection between spin and charge suggests a new method for detecting charge properties, namely via measuring spin properties.

The nitrogen vacancy (NV) center in diamond is a prototypical optically active spin qubit~\cite{Doherty_nitrogen-vacancy_2013} and has enabled advances in quantum sensing~\cite{Barry_sensitivity_2020}, where the spin state is correlated to a sensing target, such as a magnetic field. While often employed to sense phenomena external to the diamond host material, the coherence of the NV center spin is also sensitive to the behavior of nearby spins and defects within the host~\cite{Barry_sensitivity_2020}. A dominant dark spin defect in diamond is associated with the substitutional nitrogen (\ch{N_s}) center~\cite{Smith_electron-spin_1959}, which can occupy either the spin-$1/2$, neutrally charged state (\ch{N_s^0}, also termed in ESR literature the ``P1'' center, the name we use when referring to only the electron spin) or the spin-0, ionized state (\ch{N_s^+})~\cite{Lawson_existence_1998,Heremans_generation_2009,Bassett_electrical_2011} (the negatively charged state is short-lived~\cite{Ulbricht_single_2011}). The proximity of NV and \ch{N_s} centers in diamond may lead to strong dipolar interactions resolvable above background decoherence with sufficient sample design, such as through engineering the spin bath dimensionality~\cite{Marcks_guiding_2023}. This makes the system a prime candidate to study the spin-based readout of charge dynamics in a technologically relevant semiconductor.

In this work we develop a spin probe of the charge dynamics on single dark \ch{N_s} centers, measuring charge population and non-equilibrium dynamics through resonant detection of the associated spin. We initially probe the \ch{N_s} charge state through optically detected magnetic resonance (ODMR) and coherence measurements of a single nearby NV center, observing an unexpectedly low average \ch{N_s^0} neutral charge state population. We then demonstrate a spin pump-probe scheme to measure a single \ch{N_s} center spin in the presence of external stimuli. In particular, we use the \ch{N_s} spin polarization as a probe of $\ch{N_s^0}\to\ch{N_s^+}$ ionization and recapture, dynamical processes that we cannot access through traditional quantum sensing schemes based on NV center coherence. These studies demonstrate a new sensing probe of charge dynamics in semiconductors and provide practical results necessary for developing defect-based quantum devices.

\section{Results}

\begin{figure}
    \centering
    \includegraphics{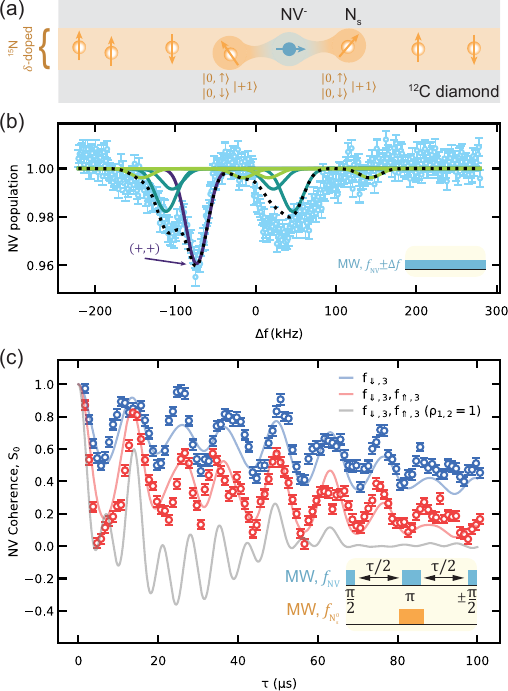}
    \caption{\textbf{Strongly coupled spin-charge defects.} (a) The studied \ch{NV^-} and \ch{N_s} centers are hosted in a two-dimensional \ch{^{15}N} $\delta$-doped layer within isotopically purified PECVD-grown diamond, a sample geometry predicted to host a higher incidence of strongly coupled electron spin systems~\cite{Marcks_guiding_2023}. (b) Optically detected magnetic resonance (ODMR, sequence inset) of the \ch{NV} center reveals strong coupling to both the \ch{N_s} spin (dipolar splitting) and charge (Stark shift) degrees of freedom. The overlaid fit arises from two \ch{N_s} centers in the $(+,+)$ (purple),  $(0/+,+/0)$ (teal), and $(0,0)$ (green) charge state combinations. (c) Time-resolved double electron-electron resonance (DEER, sequence inset) with one ($f_{\Downarrow,3}$) and two ($f_{\Uparrow,3}$ and $f_{\Downarrow,3}$) tone \ch{N_s} drives, demonstrating reduced charge populations for \ch{N_s} centers, tabulated in Table~\ref{tab:values}. Solid curves are calculated coherence for the best-matched configuration. The theoretical signal for $\rho_{1,2}=1$ is shown in grey.}
    \label{fig:characterization}
\end{figure}

\begin{table}
\caption{\label{tab:values} \bf \ch{N_s} center characterization. \rm Charge populations $\rho$, Stark shifts $d$, and dipolar couplings $a$ for each \ch{N_s} center defect.}
\begin{tabular}{|c|c|c|c|}
\hline
\ch{N_s} index  &   $\rho$    &   $d$\,(kHz)   &   $a$\,(kHz)\\
\hline
1   &   $0.474(2)$    &   $-41(1)$   &   $158.6(4)$\\
2   &   $0.302(2)$    &   $-33(2)$   &   $125(2)$\\
\hline
\end{tabular}
\end{table}

\subsection{Identifying single spin-charge defects\label{subsec:characterization}}

We study a room temperature, strongly coupled NV-\ch{N_s} system, shown schematically in Fig.~\ref{fig:characterization}(a), through which we characterize the static \ch{N_s} charge population and the spin and charge couplings to the NV center. We consider three spin-charge states of the \ch{N_s} centers, $\{\ket{0,\uparrow},\ket{0,\downarrow},\ket{+1}\}$, where the number indicates the charge state and the arrow (if applicable) denotes the spin-$1/2$ projection. The spin states couple to the NV center through dipolar interactions $a$ and the $+1$ charge state couples to the NV center through a dc Stark shift $d$~\cite{van_oort_electric-field-induced_1990,Mittiga_imaging_2018,candido_interplay_2023}. In Fig.~\ref{fig:characterization}(b) and (c) we identify both the static charge population $\rho$ of two \ch{N_s} centers and their dc Stark shift coupling to the NV center (details below), where $\rho$ is the neutral charge state population fraction. We find $\rho_1=0.474(2)$ and $\rho_2=0.302(2)$, with Stark shifts of $d_1=\SI{-41\pm1}{\kilo\hertz}$ and $d_2=\SI{-33\pm2}{\kilo\hertz}$, tabulated in Table~\ref{tab:values}, where 1 and 2 refer to the index of the \ch{N_s} center ordered by dipolar coupling strength.

Fig.~\ref{fig:characterization}(b) shows the ODMR spectrum of the NV center spin, revealing a complex structure with overlapping peaks arising from both dipolar couplings to and Stark shifts from strongly coupled \ch{N_s} centers in both charge states~\cite{van_oort_electric-field-induced_1990,Mittiga_imaging_2018,candido_interplay_2023}. This complexity motivates the use of a time-resolved double electron-electron resonance (DEER) probe that echoes out the Stark shifts, retains information about the dipolar coupling, and converts the charge population into an effective signal contrast.

In the presence of a single strongly coupled spin, where the NV-P1 dipolar coupling $a$ is larger than the background decoherence rate $\Gamma$~\cite{Marcks_guiding_2023}, the NV center coherence takes the form $S(\tau)\equiv\left<S_0\right>+i\left<S_{\pi/2}\right>=e^{-(\Gamma\tau)^n}(\cos\frac{a}{2}\tau + i \cdot p \cdot \sin\frac{a}{2}\tau)$, derived in SI Sec.~S5.B, where $n$ is a stretch factor and $p$ is the net spin polarization of the strongly coupled spin. A non-zero $p$ thus generates an out-of-phase coherence signal. In the presence of multiple strongly coupled spins the resulting coherence is the product of the signals from each single spin. Without loss of generality, here we consider only the in-phase coherence in an unpolarized spin bath $S_0(\tau)=\text{Re}[S(\tau)]$. We account for a non-unity average charge population $\rho$ of the strongly coupled defect center by modifying the coherence function (see SI Sec.~S5.C) to
\begin{equation}
	\label{eq:DEERfit}
	S_0(\tau)=e^{-(\Gamma\tau)^n} \prod_i \left[(1-\eta\cdot\rho_i)+\eta\cdot\rho_i\cdot\cos\frac{a_i}{2}\,\tau\right]
\end{equation}
where $\eta$ captures the fraction of the \ch{N_s} spin population addressed ($\eta=3/8$ or $6/8$ for one and two-tone DEER measurements, respectively, described in the next paragraph), and $\rho_i$ and $a_i$ are the neutral charge population fraction and dipolar spin coupling for the $i^{th}$ \ch{N_s} center.

All measurements are performed under an external magnetic field of \SI{412}{G} aligned within \SI{0.5}{\degree} of the [111] diamond crystal axis (also the NV quantization axis). Under this condition there are four non-degenerate P1 electron spin transitions, corresponding to the two \ch{^{15}N} nuclear spin states and combinations of the four electron Jahn-Teller (JT) states~\cite{Smith_electron-spin_1959}. We refer to these four frequencies as $f_{\Downarrow,1}$, $f_{\Downarrow,3}$, $f_{\Uparrow,3}$, and $f_{\Uparrow,1}$, where the arrow refers to the nuclear spin state and the number refers to number of degenerate JT state addressed. The JT-dependent hyperfine coupling is sufficient to spectrally isolate the four transitions. We measure the NV center coherence under the DEER scheme (with a pulse sequence illustrated by the inset of Fig.~\ref{fig:characterization}(c)), driving $\pi$-pulses at either one ($f_{\Downarrow,3}$) or two ($f_{\Downarrow,3}$ and $f_{\Uparrow,3}$) resonance tones.

We reconstruct the configuration of \ch{N_s} centers around the NV center from the DEER measurement described above in order to extract the \ch{N_s} center charge properties. We calculate the DEER coherence signal for $10^7$ statistically random configurations of \ch{N_s} centers, accounting for $\rho<1$. We recover as the best fit a configuration containing two strongly coupled \ch{N_s} centers with dipolar couplings and non-unity average neutral charge populations listed in Table~\ref{tab:values}. These values are then held constant to fit the spectrum in Fig.~\ref{fig:characterization}(b) to extract the Stark shifts from the ionized \ch{N_s^+} centers. The details of the DEER fit procedure and the effect of the interactions between two $\ch{N_s}$ centers with $\rho<1$ are discussed in SI Sec. S2. The simulated coherence for this configuration is plotted in the solid lines atop the data in Fig.~\ref{fig:characterization}(c) for the one- and two-tone measurements. The grey solid line plots the calculated two-tone coherence fixing $\rho=1$ for all \ch{N_s} centers, demonstrating that the NV coherence contrast is reduced due to $\rho<1$~\cite{dwyer_probing_2022}.

\subsection{Dark spin polarization and readout\label{subsec:P1polread}}

\begin{figure*}
    \centering
    \includegraphics[width=\textwidth]{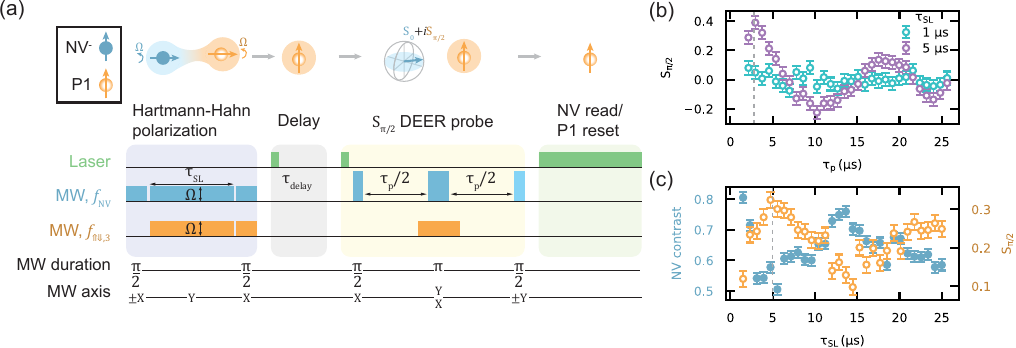}
    \caption{\textbf{Dark spin polarization and measurement.} (a) NV-P1 polarization pump-probe measurement sequence. Hartmann-Hahn spin locking at Rabi rate $\Omega$ turns on polarization transfer between the initialized NV electron spin and the mixed P1 electron spin. Initial NV pulse axis ($\pm X$) determines the sign of the P1 polarization, enabling differential measurement of the P1 spin population. The P1 spin polarization (signified by the shaded circle) evolves during an arbitrary delay time. An out-of-phase DEER measurement, $S_{\pi/2}$, on the NV probes the P1 polarization. The NV is driven at $2\Omega$ in this segment to avoid resonance with the P1 during the $\pi$-pulse. After NV readout, a long laser pulse resets the P1 spin state, explored more completely in Sec.~\ref{subsec:laserdecay}. When overlapping microwave pulses are present, the duration refers to both pulses, with duration referenced to the drive period. When only one microwave axis is present, it refers to both pulses. (b) $S_{\pi/2}$ versus probe time $\tau_p$ for two different fixed spin locking times. (c) NV polarization, measured optically, after a variable spin locking drive $\tau_{SL}$ (blue, filled circles) and P1 polarization, measured through $S_{\pi/2}$ (orange, open circles), after the same time. Correspondence between the two indicates that $S_{\pi/2}$ properly measures the polarization transferred to the P1 spin. $\tau_p$ and $\tau_{SL}$ times that yield the largest signals are marked in (b) and (c), respectively, with dashed grey lines.}
    \label{fig:P1readout}
\end{figure*}

The above DEER analysis motivates the use of the P1 spin polarization $p$ as a precise probe of \ch{N_s} charge dynamics. Extending previously demonstrated techniques~\cite{Knowles_demonstration_2016,Rosenfeld_sensing_2018,Cooper_identification_2020}, a pump-probe measurement of the strongly coupled P1 center spin polarization is achieved via the combination of resonant and off-resonant dipolar interactions between the NV and P1 electron spins shown in Fig.~\ref{fig:P1readout}(a), with schematics of the relevant interactions shown above the pulse sequence. This method constructs a spin-spin-photon interface between the dark P1 center spin and the optically active NV center spin, providing effective optical readout of the P1 center spin polarization. Throughout the measurements in this section and Sec.~\ref{subsec:laserdecay} we use two microwave tones at $f_{\Downarrow,3}$ and $f_{\Uparrow,3}$ to enhance $\eta$ and improve signal collection. 

After optical initialization of the NV center into the $\ket{m_s=0}$ state, a spin lock-driven Hartmann-Hahn resonance condition (first panel) activates $\h S_+\h P_-+h.c.$ dipolar spin exchange interactions, where $\h S$ is the NV electron spin operator and $\h P$ is the P1 electron spin operator, between the NV and all driven P1 spins. In a strongly coupled spin system, coherent polarization transfer to and from the nearest P1 spin dominates background polarization diffusion. After driving a Hartmann-Hahn resonance for time $\tau_{SL}$ the NV spin state is pumped into the P1 spin, and the NV is repolarized with a laser pulse short enough to avoid perturbing the P1 spin state significantly (see Sec.~\ref{subsec:laserdecay}). The P1 spin is then allowed to freely evolve during a delay period (see second panel of Fig.~\ref{fig:P1readout}(a)). Additional stimuli may be added here, such as microwave and laser pulses, to modify the P1 or NV evolution.

An out-of-phase DEER measurement (third panel in Fig.~\ref{fig:P1readout}(a)) of the NV coherence, $S_{\pi/2}(\tau)=\text{Im}[S(\tau)]$, directly probes the P1 spin polarization, as evidenced by Eq.~\eqref{eq:DEERfit}. In this segment the NV drive amplitude is doubled to avoid any resonant interactions with the P1 during the $\pi$-pulse. Details of this readout interaction can be found in SI Sec.~S4 and S5. After the P1 readout, both the NV and the P1 are reset with a sufficiently long laser pulse  $\approx\SI{100}{\micro\second}$ (fourth panel, see Sec.~\ref{subsec:laserdecay} for details).

Because of the low extracted charge population, simultaneous occurrence of the two \ch{N_s} centers in the neutrally charged spin-1/2 state is statistically negligible compared to the case of one neutrally charged center and one ionized center (see SI Sec.~S5.C for detail). Given the small separation between $a_1$ and $a_2$, we may tune the polarization and probe steps of the measurement to maximize signal and assume we average over the four possible charge combinations, while remaining sensitive to both spins. 

In Fig.~\ref{fig:P1readout}(b), $S_{\pi/2}$ is measured versus DEER probe time $\tau_p$ at two different spin lock times, $\tau_{SL}=1, \SI{5}{\micro\second}$. At $\tau_{SL}=\SI{5}{\micro\second}$, corresponding to $\approx 1/(4a_{1,2})$, we observe oscillations that match the strong coupling observed in Fig.~\ref{fig:characterization}(c), as expected. We extract $\tau_p=\SI{2.5}{\micro\second}$ as the probe point that maximizes readout signal, marked with a dashed grey line (the x-axis is shifted by the $\pi$-pulse time on the NV center).

Fig.~\ref{fig:P1readout}(d) demonstrates the direct probe of coherent polarization exchange between the NV and P1 electron spins. The blue circles show the NV population measured after Hartmann-Hahn polarization transfer (the leftmost panel in Fig.~\ref{fig:P1readout}(a)) at drive rate $\Omega\approx\SI{400}{\kilo\hertz}$, revealing resolvable coupling to neighboring P1 spins. The orange open circles show the P1 spin polarization read through $S_{\pi/2}(\tau_p=\SI{2.5}{\micro\second})$ for the same $\tau_{SL}$, with no delay ($\tau_{delay}=\SI{0}{\micro\second}$). We observe that as polarization leaves the NV spin, it transfers to the P1 spin. Probing at $\tau_p=\SI{2.5}{\micro\second}$ ensures that $S_{\pi/2}$ is only sensitive to polarization of the nearby P1 spin. We identify $\tau_{SL}=\SI{5}{\micro\second}$, marked with a dashed grey line, as the optimal spin locking time to maximize polarization transfer.

In comparison to NV correlation measurements that have probed spin bath evolution, for example in diamond surface spins~\cite{sushkov_magnetic_2014,Rezai_probing_2022,dwyer_probing_2022}, the resolution of \ch{N_s} center dynamics is not limited by the NV center $T_1$ lifetime, as no signal is stored in the NV spin state. As long as the laser power is low enough such that the NV repolarization pulse does not disturb the P1 electron spin, the measurement is only limited by shot noise.

\subsection{Non-equilibrium charge dynamics\label{subsec:laserdecay}}

\begin{figure}
    \centering
    \includegraphics{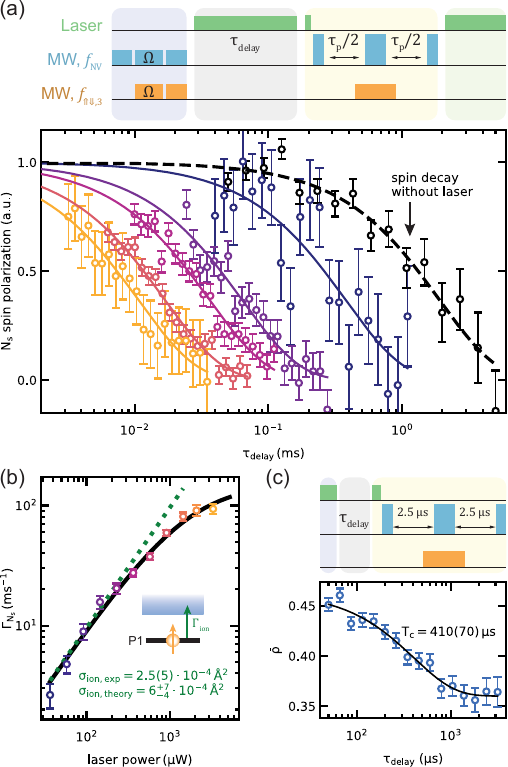}
    \caption{\textbf{Ionization-driven spin decay.} (a) Top, pulse sequence for time-resolved P1 spin decay under green illumination. After P1 spin initialization, the laser is turned on for variable time $\tau_{delay}$, after which the $S_{\pi/2}$ probe is performed. The reset laser pulse at the end is fixed at a power-dependent time long enough to fully decay the spin polarization. Bottom, P1 polarization decay under range of laser powers (color-coded in (b)), with mono-exponential fits overlaid. The right-most, black curve measures the P1 spin lifetime in the dark, without the $\tau_{delay}$ laser pulse. (b) $\Gamma_{\textrm{N}_s}$ decay rates extracted from (a) versus laser power, with saturation fit in black and low-power linear fit in dashed green. Extracted ionization cross-section is indicated. (c) Decay of average charge population $\bar{\rho}$ readout through NV center $S_0$ coherence contrast, pulse sequence shown at top.}
    \label{fig:laserdecay}
\end{figure}

In the previous section we combined on- and off-resonant microwave controls to realize a spin-spin-photon interface between the P1 center and NV center electron spins. The demonstrated technique provides a spin-based optical measurement to monitor the evolution of the \ch{N_s} charge, which we now utilize to measure ionization under \SI{2.33}{\eV} laser illumination. This laser energy is common for NV center magnetometry, and is large enough to ionize the \ch{N_s^0} electron, which sits $\SIrange{2.1}{2.3}{\eV}$ below the diamond conduction band (see SI Sec. S10). The pertinent components of the measurement are shown in the sequence in Fig.~\ref{fig:laserdecay}(a), with a variable-power illumination during the delay window. We first measure the P1 center spin polarization decay without illumination (i.e., the laser is off during the delay window). The results are plotted in Fig.~\ref{fig:laserdecay}(a), where we find a spin $T_1=\SI{1.9\pm.2}{\milli\second}$, in agreement with prior bulk EPR measurements of \ch{^{14}N} P1 ensemble spin relaxation that yield room temperature phonon-limited relaxation times $\approx \SIrange{1}{3}{\milli\second}$~\cite{Reynhardt_temperature_1998} (see SI Sec.~S8).

Figs.~\ref{fig:laserdecay}(a) and (b) present the \ch{N_s^0}/P1 spin polarization decay and mono-exponential decay rates $\Gamma_{\textrm{N}_s}$ as a function of laser power $P$ ranging from \SIrange{36}{3300}{\micro\watt}. All the observed decay times are at least an order of magnitude longer than the NV spin repolarization time, which ranges from \SIrange{2.8}{0.24}{\micro\second} in the measured power range. This separation of time-scales is critical, leaving the P1 spin polarization undisturbed during the short repolarization laser pulse (see SI Sec.~S10). We fit $\Gamma_{\textrm{N}_s}$ empirically to a saturation curve $\Gamma_{\textrm{N}_s}(P)=\Gamma_{sat}\frac{P}{P+P_{sat}}$ for rate $\Gamma_{sat}$ at saturation power $P_{sat}$. We find saturation at $P_{sat}=\SI{1.6\pm.3}{\milli\watt}$, likely arising from saturation of surrounding charge traps~\cite{Grivickas_carrier_2020,shinei_change_2023}.

At low laser powers we extract an ionization cross section of $\sigma_{ion,exp}=\SI{2.5\pm.5 e-4}{\AA^2}$, similar to prior work in ensemble measurements~\cite{Dhomkar_charge_2018,Isberg_photoionization_2006} (see SI Sec.~S9 for analysis details). We corroborate this value with density functional theory (DFT) calculations of the ionization cross-section $\sigma_{ion,theory}$ of the \ch{N_s^0} center, detailed in SI Sec.~S11. The experimental cross section falls within the calculated range $\sigma_{ion,theory}=6^{+7}_{-4}\times 10^{-4}\,\textrm{\AA}^2$, where the range of values arises from uncertainty in the \ch{N_s^0} energy level.

These results represent the first observation of coupled spin and charge dynamics in an optically inactive defect center with single-defect resolution. Furthermore, we demonstrate that the P1 spin polarization persists under ionizing radiation, complementing previous work that has studied \ch{N_s^0} ionization through charge transport~\cite{Heremans_generation_2009} and Stark shifts on the NV center spin~\cite{Bassett_electrical_2011}.

We further probe the charge dynamics with a time-resolved charge decay measurement~\cite{Lozovoi_dark_2020}, using the pulse sequence shown on the top of Fig.~\ref{fig:laserdecay}(c). After a sufficiently long laser pulse to reset the \ch{N_s} charge states, the system is allowed to evolve in the dark for a variable time. We then probe $S_0$ at the first coherence dip in Fig.~\ref{fig:characterization}(c), which gives us access to the geometric average charge population $\bar\rho=\sqrt{\rho_1\rho_2}$. We observe a signal corresponding to an initial average population $\bar{\rho}(0)=0.463(8)$ that decays to a steady-state value of $\bar{\rho}=0.360(6)$, close to the value extracted from Fig.~\ref{fig:characterization}(d) of $0.378(2)$, with exponential decay time $T_c=\SI{410\pm70}{\micro\second}$.

\begin{figure}
	\centering
	\includegraphics{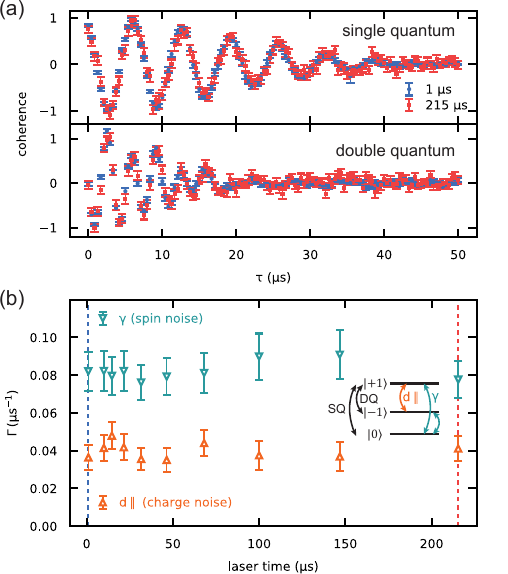}
	\caption{\bf Spin and charge noise. \rm (a) Single (SQ, top) and double (DQ, bottom) quantum Ramsey measurements on a second NV center with two different laser initialization times. (b) Electric ($d\parallel$) and magnetic ($\gamma$) field noise, corresponding to charge and spin density, respectively, extracted from SQ and DQ Ramsey decay rates for varying laser initialization times.}
	\label{fig:laser-coherence}
\end{figure}

Quantum sensing schemes often rely on detecting changes in spin polarization or coherence lifetimes of a central spin. Fig.~\ref{fig:laser-coherence} explores if the charge dynamics observed in Fig.~\ref{fig:laserdecay} are detectable through standard coherence measurements on a second NV with no strongly coupled \ch{N_s} centers. Fig.~\ref{fig:laser-coherence}(a) shows the single quantum (SQ, top) and double quantum (DQ, bottom) Ramsey decay~\cite{candido_interplay_2023} with two laser initialization times, chosen based on Fig.~\ref{fig:laserdecay} to be long (short) enough to fully (not) ionize the \ch{N_s} electron. Fig.~\ref{fig:laser-coherence}(b) extracts the environmental magnetic field/spin noise and electric field/charge noise from the SQ and DQ measurements~\cite{candido_interplay_2023}. We find no change in decay rate, and thus spin and charge densities, as the laser time is increased, indicating no modulation of spin or charge densities after ionization in this sample. This lies in contrast to other recent observations, which were performed on a sample grown under different conditions and likely containing a different crystal defect environment and density than in the one studied here~\cite{Wang_manipulating_2023}.

\section{Discussion}
Leveraging the strong coupling between optically active and dark spin defects, we implement a spin pump-probe measurement for in-situ charge sensing. Specifically, we measure static and dynamic charge processes in a semiconductor via the intrinsic correlation between defect charge and spin states. The non-unity charge population of single \ch{N_s} centers in diamond has not been previously observed, and measuring the ionization and recapture of electrons generally requires ultrafast optical spectroscopy~\cite{Ulbricht_single_2011}. Our spin-based technique relies only on modifications to the standard spin resonance and coherence measurements from quantum sensing and ESR~\cite{Barry_sensitivity_2020}, making it applicable in many semiconductor systems and circumventing the need for ultrafast pulsed lasers.

Our results suggest new avenues in quantum sensing and the development of new qubit host materials. Our technique may be generalized to other quantum sensing platforms~\cite{Gottscholl_initialization_2020,Bayliss_optically_2020} to study charge dynamics in a range of materials, including both the qubit host materials and external systems, such as molecular systems and transition metal dichalcogenides~\cite{Regan_emerging_2022}. It also provides a precise method to characterize qubit host materials at a level relevant for the development of quantum technologies. Other platforms based on defects in semiconductors, such as the recently demonstrated \ch{Er^{3+}:CeO2}~\cite{Zhang_optical_2023}, will benefit from the in-situ probe of the defect's local charge environment.

The spin probe of charge fluctuations may be applied to the study of noise sources on the surface of diamond~\cite{dwyer_probing_2022,Rezai_probing_2022,zhang_nanoscale_2021,Xie_biocompatible_2022}, which can arise from both magnetic and electric field fluctuations from near-surface or surface-residing particles~\cite{dwyer_probing_2022,candido_interplay_2023}. A reasonable regime for studying single strongly coupled surface spins is discussed in SI Sec.~S12, which would enable a direct measure of these noise sources.

\section{Methods}
\subsection{Sample fabrication}
The NV and \ch{N_s} centers are all hosted within a roughly \SI{4}{\nano\meter}, few ppm nitrogen $\delta$-doped~\cite{Ohno_engineering_2012} layer \SI{50}{\nano\meter} below the diamond surface, within \ch{^{12}C} isotopically purified [100] diamond, grown at Argonne National Laboratory via plasma-enhanced chemical vapor deposition (PE-CVD). Details and characterization of the PE-CVD can be found in Ref.~\cite{Marcks_guiding_2023}. NV center synthesis is achieved via vacancy creation with \SI{2e14}{\centi\meter^{-2}}, \SI{2}{\mega\eV} electron irradiation followed by a \SI{2}{\hour}, $850^\circ$C anneal under Ar atmosphere.

\subsection{Spin measurements}
Optical measurements are performed with a \SI{532}{\nano\meter} Oxxius LCX-532S-150-CSB-PPA CW laser amplitude-modulated by a custom Gooch \& Housego acousto-optic modulator (AOM) and focused through an Olympus MPLFLN100x 0.9NA objective. The laser power is adjusted with variable optical density filters. Photon counting is performed with a Perkin Elmer SPCM-AQRH-13 avalanche photodiode (APD) connected to a NI DAQ USB-6341 through Minicircuits ZYSWA-2-50DR switches.

Spin control of the NV center is performed with a SRS SG396 signal generator. Spin control of the P1 center is performed with a SRS SG384 at $f_{\Downarrow,3}$ and a HMC1197 eval board at $f_{\Uparrow,3}$ signal generators, combined with a Minicircuits ZFRSC-123-S+ power combiner. Microwave signals are gated with Minicircuits ZASWA-2-50DR+ switches and amplified with a Minicircuits ZX60-83LN12+ low noise amplifier followed by a Minicircuits ZHL-16W-43+ high power amplifier.

Experiment timing is controlled with a Swabian Pulsestreamer 8/2 AWG and experiments are performed in nspyre experimental physics software~\cite{feder_nspyre_2023}.

\section{Acknowledgements}
The authors would like to thank Grant T. Smith, Paul Jerger, and Denis R. Candido for helpful discussions.

This work was primarily supported by the U.S. Department of Energy, Office of Science, Basic Energy Sciences, Materials Sciences and Engineering Division (J.C.M., F.J.H., D.D.A.), with additional support from Midwest Integrated Center for Computational Materials (MICCoM) as part of the Computational Materials Sciences Program funded by the US Department of Energy (M.O. and G. G.), the Q-NEXT Quantum Center as part of the US Department of Energy, Office of Science, National Quantum Information Science Research Centers (N.D., M.F.), the Center for Novel Pathways to Quantum Coherence in Materials, an Energy Frontier Research Center funded by the US Department of Energy, Office of Science, Basic Energy Sciences (Y.-X. W.), and the AFOSR program FA9550-22-1-0370 Engineering Targeted Quantum Characteristics in SiC-based Materials (Y.J., Y.Z., M.O.).  J.C.M. acknowledges prior support from the National Science Foundation Graduate Research Fellowship Program (grant no. DGE-1746045). M.O. acknowledges the support from the Google PhD Fellowship. This work made use of the Pritzker Nanofabrication Facility of the Pritzker School of Molecular Engineering at the University of Chicago, which receives support from Soft and Hybrid Nanotechnology Experimental (SHyNE) Resource (NSF ECCS-2025633), a node of the National Science Foundation's National Nanotechnology Coordinated Infrastructure. This research also used resources of the National Energy Research Scientific Computing Center, a DOE Office of Science User Facility supported by the Office of Science of the U.S. Department of Energy under Contract No.~DE-AC02-05CH11231 using NERSC award ALCC-ERCAP0025950, and resources of the University of Chicago Research Computing Center.

\section{Author contributions}
J.C.M. conceived of the study, carried out measurements, analyzed data, and wrote the initial manuscript. M.O. performed coherence calculations and analyzed data. M.O., Y-X.W., and M.F. provided theoretical insight. B.S. assisted with measurements. Y.J. and Y.Z. performed ionization calculations. N.D. grew the diamond sample. F.J.H., A.A.C., G.G., and D.D.A. obtained funding and oversaw project. All authors contributed to manuscript preparation.

\section{Competing Interests}
The authors declare no competing interests.

%

\end{document}